\documentclass[aps,prd,twocolumn,superscriptaddress,amsfont,graphicx,nofootinbib,preprintnumbers]{revtex4}%
\usepackage{color,graphicx,epsfig}
\usepackage{ifpdf}
\usepackage{amsmath}
\usepackage{bm}
\usepackage{color}
\usepackage[english]{babel}
\usepackage{graphicx}%
\usepackage{amsfonts}%
\usepackage{amssymb}
\usepackage{braket}
\usepackage{hyperref}

\definecolor{nicered}{rgb}{0.7,0.1,0.1}
\definecolor{nicegreen}{rgb}{0.1,0.5,0.1}
\hypersetup{colorlinks,citecolor= nicegreen,linkcolor= nicered}

\newcommand{\beq}{\begin{equation}}
\newcommand{\eeq}{\end{equation}}
\newcommand{\bea}{\begin{eqnarray}}
\newcommand{\eea}{\end{eqnarray}}

\DeclareMathOperator{\Li}{Li}
\DeclareMathOperator{\HPL}{H}

\begin{document}

\def\LjubljanaFMF{Faculty of Mathematics and Physics, University of Ljubljana,
 Jadranska 19, 1000 Ljubljana, Slovenia }
\def\LjubljanaIJS{Jo\v zef Stefan Institute, Jamova 39, 1000 Ljubljana, Slovenia}
\def\CERN{CERN, Theory Division, CH-1211 Geneva 23, Switzerland}
\def\Oxford{Rudolf Peierls Centre for Theoretical Physics, University of Oxford
OX1 3PN Oxford, United Kingdom}

\title{Searching for new spin-0 resonances at LHCb}

\author{Ulrich Haisch} 
\email[Electronic address:]{Ulrich.Haisch@physics.ox.ac.uk} 
\affiliation{\Oxford}
\affiliation{\CERN}

\author{Jernej F.\ Kamenik} 
\email[Electronic address:]{jernej.kamenik@cern.ch} 
\affiliation{\CERN}
\affiliation{\LjubljanaIJS}
\affiliation{\LjubljanaFMF}

\preprint{CERN-TH-2016-010}

\date{\today}
\begin{abstract}
We study the phenomenology of light spin-0 particles and stress that they can be efficiently searched for at the LHCb experiment in the form of dimuon resonances. Given the  large production cross sections in the forward rapidity region together with the efficient triggering and excellent mass resolution, it is argued that LHCb can provide unique sensitivity to such states. We illustrate our proposal using the recent measurement of Upsilon production by LHCb, emphasising the importance of mixing effects in the bottomonium resonance region. The implications for  dimuon decays of spin-0 bottomonium states are also briefly discussed. 
\end{abstract}

\maketitle

\section{Introduction}

The presence of scalar particles is known to be tightly related to the phenomenon of symmetry breaking. The role of scalar degrees of freedom in fundamental theories has therefore been of lasting interest in both experimental and theoretical physics. These efforts have been refocused with the discovery of the Higgs boson, whose measured properties \cite{Aad:2015gba,Khachatryan:2014jba}  suggest that it provides the dominant source of breaking of both the electroweak (EW) and flavour symmetries of the standard model (SM).  

Another general property of Higgs-like fields is that they can act as portals \cite{Silveira:1985rk,McDonald:1993ex,Burgess:2000yq} between the SM with its gauge and accidental symmetries and other hypothetical particles, that are neutral under the SM symmetries --- so-called dark sectors, possibly including dark matter. These features provide a strong motivation for continuing searches for new scalar degrees of freedom both at high and low energies. 

New scalars coupling to the SM fermions necessarily carry SM flavour quantum numbers or conversely break the SM flavour symmetry. The  overall agreement of most measured flavour observables with the corresponding SM predictions however severely restricts any new source of  flavour breaking. The simplest, but also most restrictive solution to this problem is to assume that also beyond the SM, the minimal possible flavour breaking consistent with the observed fermion mass and mixing patterns  is realised. This assumption often goes under the name of minimal flavour violation~(MFV)~\cite{D'Ambrosio:2002ex}. It leads to the clear prediction that the couplings between any new neutral spin-0 state and SM matter are predominantly flavour conserving and proportional to the fermion masses. 

Even beyond MFV, simply requiring agreement with the existing constraints on new fermion interactions coming from precision low-energy (mostly flavour) experiments typically leads to severe restrictions on the size of flavour-violating and CP-violating couplings. In particular, couplings to lighter fermion generations have to be highly suppressed, while the couplings to heavier quarks and leptons are less constrained. One is thus naturally led to consider new  spin-0 particles that couple most strongly to the  third generation. Similar to the SM Higgs, such resonances tend to decay to the heaviest kinematically allowed  final state and can be abundantly produced in hadronic high-energy  collisions through loop-induced gluon-gluon fusion, provided they couple to quarks and are sufficiently light.    

Despite their potentially large production cross sections, it however turns out that new third-generation-philic spin-0 particles may have escaped detection in existing experiments even for  moderately large couplings, if they have masses in the ballpark of $[10, 50] \, {\rm GeV}$. First of all, given their very small couplings to electrons and EW  gauge bosons such states  easily pass most LEP constraints (see however~\cite{Abbiendi:2001kp}). Only for masses below about $10 \, {\rm GeV}$ do radiative Upsilon  $\big($$\Upsilon (n)$$\big)$ decays~\cite{Lees:2011wb,Lees:2012iw,Lees:2012te} and rare $B$-meson decays~\cite{Schmidt-Hoberg:2013hba, Dolan:2014ska, Aaij:2015tna}  provide stringent constraints. While masses above $50 \, {\rm GeV}$ are already probed by LHC run~I data, the existing searches typically become ineffective for softer final states due to trigger requirements and loss of acceptance (counterexamples include~\cite{Chatrchyan:2012am, ATLAS:2011cea}). The situation is actually expected to worsen at the higher energies explored at  LHC run II and beyond.  In the following, we will argue that this represents a great opportunity for the LHCb experiment with its efficient triggering, excellent vertexing and accurate event reconstruction to provide unique probes of new spin-0 particles with masses in the range from few~GeV to few~tens of GeV. 

This article is structured as follows. In Sec.~\ref{sec:generalities} we discuss the  relevant  spin-0  interactions and study the resulting branching ratios and production cross sections. Our new search strategy for light  Higgs-like particles at LHCb is introduced in Sec.~\ref{sec:searchstrategy}, and applied to bottomonium states in Sec.~\ref{sec:etabdimuonBr}. The numerical analyses for the spin-0 cases are performed in Sec.~\ref{sec:spin0bounds}. We conclude in~Sec.~\ref{sec:conclusions}. Formulas for the partial widths and branching ratios of spin-0 states are collected  in~App.~\ref{app:widths}. 

\section{Generalities}
\label{sec:generalities}

We choose to work within an effective theory description below the EW breaking scale ($v \simeq 246 \, {\rm GeV}$), where the relevant Lagrangian is given by
\beq \label{eq:simplifiedmodel}
\begin{split}
\mathcal L  & =  \mathcal L_{\rm SM} + \frac{1}{2} \left[ (\partial P)^2 - m^2_P P^2 + (\partial S)^2 - m_S^2 S^2 \right] \\[1mm] & \phantom{xx} -  \sum_f \frac{m_f}{v} \left( i \kappa^f_P P \bar f \gamma_5 f +\kappa^f_S S \bar f f \right) \,,
\end{split}
\eeq
with ${\cal L}_{\rm SM}$ encoding the SM interactions. One can easily match the above interactions to more complete EW  descriptions above the weak scale, such as multi-Higgs models. In writing (\ref{eq:simplifiedmodel}), we have assumed that the new spin-0 particles $P,S$ couple to all SM fermions~$f$ in a flavour-conserving way and that their interactions conserve CP, which renders the coefficients $\kappa^f_{P,S}$ real. As already discussed in the introduction, both assumptions are phenomenologically well motivated due to the existing stringent constraints on new sources of flavour and CP violation (cf.~\cite{Freytsis:2009ct, Batell:2009jf, Kamenik:2011vy,Clarke:2013aya, Schmidt-Hoberg:2013hba,Dolan:2014ska,DMLHCtalk}). However, even the addition of small flavour off-diagonal or CP-violating couplings consistent with current constraints would neither affect our general discussion nor our conclusions. The same applies to possible couplings of the new mediators to the SM Higgs and EW gauge bosons, which if present, can provide additional constraints on such scenarios. Since these constraints are strongly  model dependent, we will not consider them in what follows. 

The simplified model  (\ref{eq:simplifiedmodel}) is valid as long as the new  scalar $S$ does not mix strongly with the SM Higgs boson and there are no additional light degrees of freedom  below the EW scale. In such a case the model dependence associated to the full Higgs sector is encoded in the portal couplings $\kappa^f_{P,S}$.  The simplest choice of couplings is universal~$\kappa^f_{P,S} = \kappa_{P,S}$ and realised in singlet scalar extensions of the SM Higgs sector. Within the decoupling limit of the two-Higgs-doublet model type II (THDMII), one has instead $\kappa^{e,\mu,\tau,d,s,b}_{P,S} = \tan\beta $, $\kappa_{P,S}^{u,c,t} = \cot \beta$ with $\tan \beta$  denoting the ratio of vacuum expectation values of the two Higgs doublets. More generally, the MFV hypothesis allows for $\kappa_{P,S}^{d,s,b} = \kappa_D$, $\kappa_{P,S}^{u,c,t} = \kappa_U$  in the quark sector and $\kappa_{P,S}^{e,\mu,\tau} = \kappa_L$ for charged leptons. 

\begin{figure}[!t]
\centering
\includegraphics[angle=0,width=8.75cm]{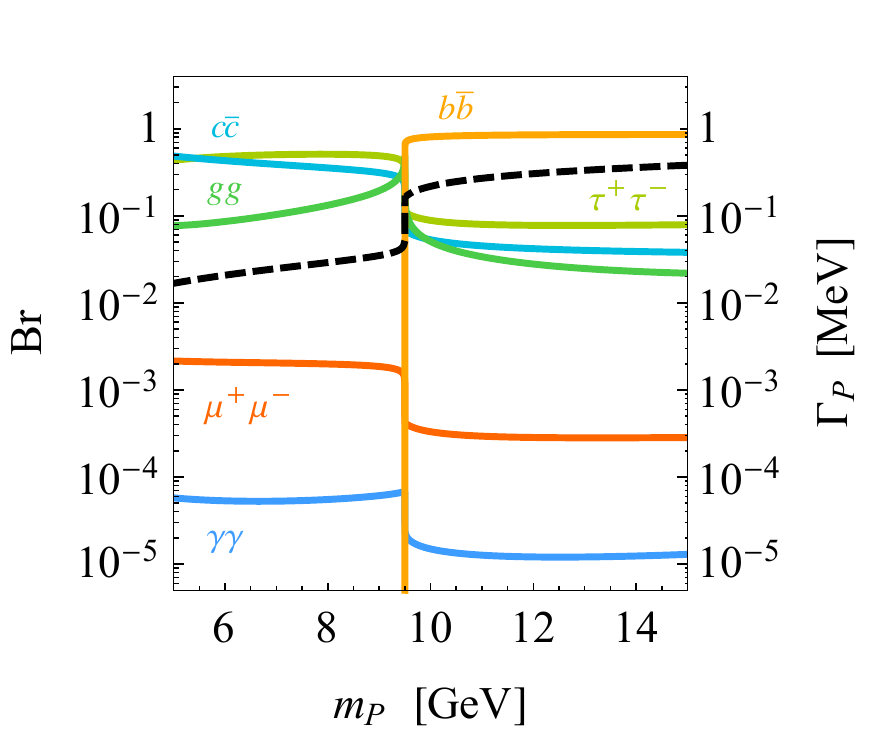}
\includegraphics[angle=0,width=8.75cm]{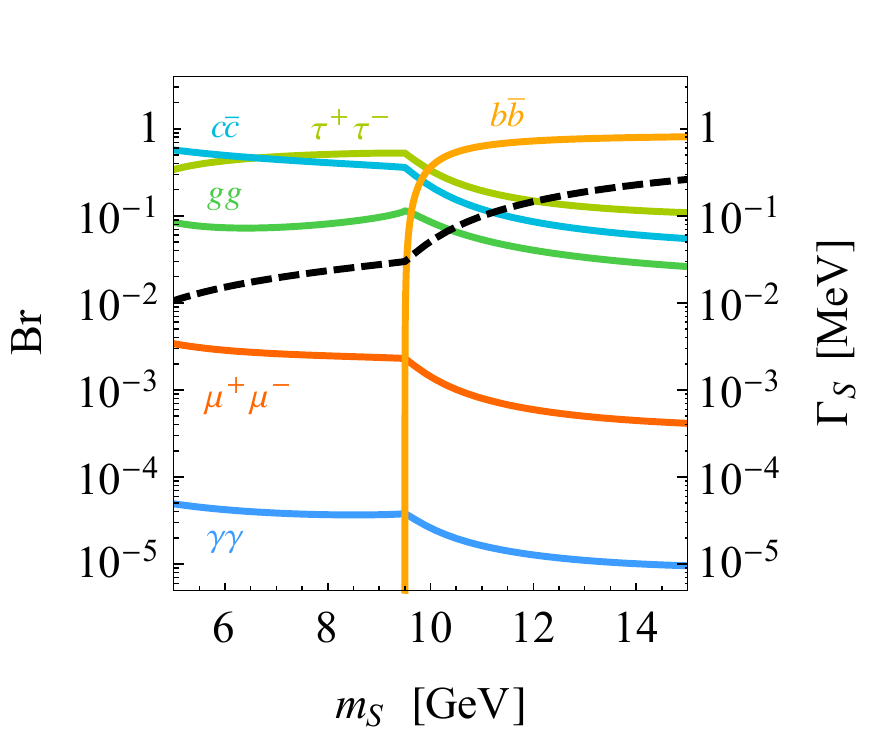}
\caption{\label{fig:BrGa} Branching ratios (coloured curves) and total decay widths (black dashed curves) of  a pseudoscalar (upper panel) and scalar (lower panel) with $\kappa_{P,S}^f =1$. Effects of  mixing and hadronisation have not been included in  these predictions.}
\end{figure}

In the  mass range  of interest and under the assumption that the couplings $\kappa^f_{P,S}$ are approximately universal, the mediators~$P,S$ decay dominantly to $b\bar b$ (for $m_{P,S} > 2 m_b$), $c\bar c$ and $\tau^+ \tau^-$. Somewhat suppressed are instead the  $\mu^+\mu^-$ and $\gamma \gamma$ branching ratios. These features are illustrated in the two panels of Fig.~\ref{fig:BrGa}. From these plots it is also evident that for $\big | \kappa_{P,S}^{f} \big | \lesssim \mathcal O(1)$ the new resonances will be very narrow with the total decay widths not exceeding~$1 \, {\rm MeV}$. The shown results are obtained using the formulas given in~App.~\ref{app:widths}. 

\begin{figure}[!t]
\centering
\includegraphics[angle=0,width=7cm]{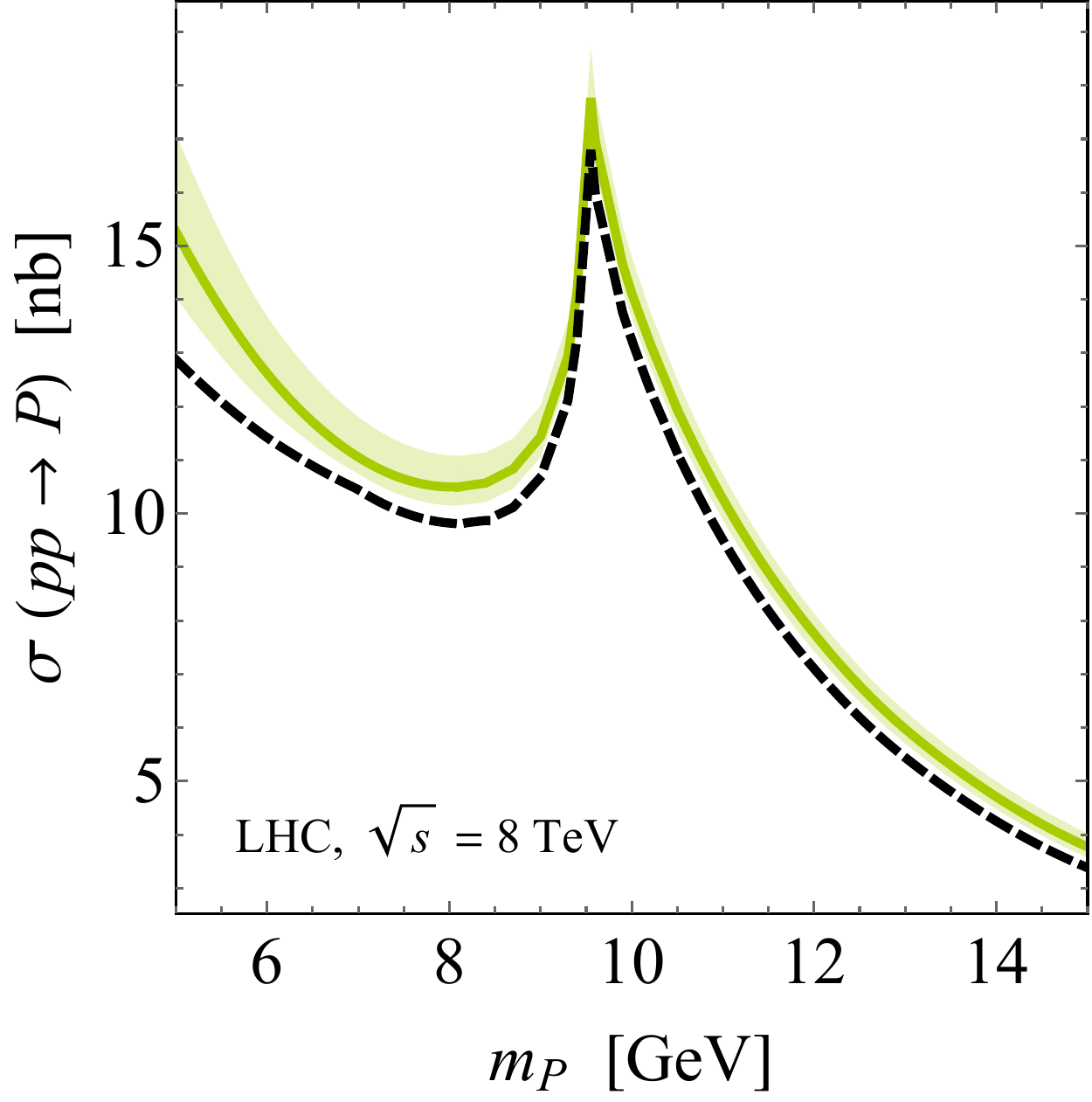}
\vskip0.5cm
\includegraphics[angle=0,width=7cm]{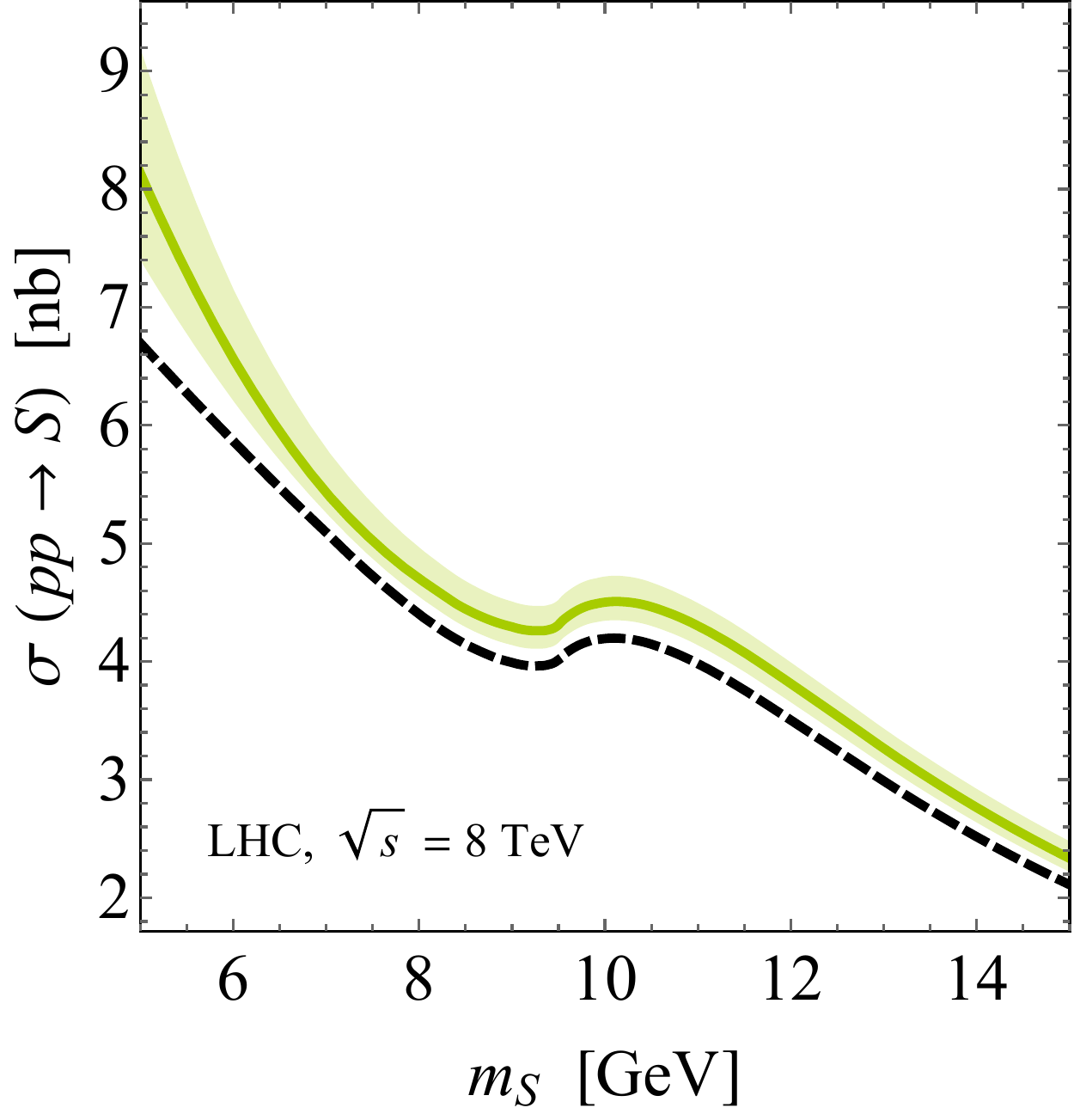}
\caption{\label{fig:xsecs} Inclusive production cross section of a pseudoscalar~(upper panel) and scalar (lower panel) at the $8 \, {\rm TeV}$ LHC, assuming $\kappa_{P,S}^f=1$. Mixing and hadronisation effects have not been included in these predictions. The cross section estimates used to set bounds on the new spin-0 resonances are shown as  black dashed curves and correspond to theoretical 95\%~CL lower limits. See text for details.}
\end{figure}

The relative suppression of the clean $\mu^+\mu^-$ and $\gamma \gamma$ final states, however turns out to be of no big concern in practice given the sizeable  production rates of light spin-0 states at the LHC. From Fig.~\ref{fig:xsecs}, one sees that the inclusive cross sections at $8 \, {\rm TeV}$  for a  scalar or pseudoscalar of ${\cal O} ( 10 \, {\rm GeV})$ mass range from a few~${\rm nb}$ to tens of ${\rm nb}$. The depicted next-to-leading order~(NLO) QCD results have been obtained with~{\tt HIGLU}~\cite{Spira:1995mt} and employ {\tt NNPDF30\_nlo\_as\_0118\_lhcb} parton distribution functions (PDFs)~\cite{Gauld:2015yia}. The use of this specific~set is motivated by the fact that these PDFs allow for a better description of small-$x$ physics, because they include besides the data incorporated in the standard {\tt NNPDF30\_nlo\_as\_0118} fit \cite{Ball:2014uwa} information on prompt charm production at LHCb~\cite{Aaij:2013mga}. The theory uncertainties  that are displayed as coloured bands in the plots of Fig.~\ref{fig:xsecs}, include both PDF and scale ambiguities. The former are obtained by calculating the~68\% confidence level (CL) envelope of all  50 members of  the {\tt  NNPDF30\_nlo\_as\_0118\_lhcb} set,  while the latter are determined by identifying  the renormalisation and factorisation scales $\mu=\mu_R=\mu_F$ and varying $\mu$ in the range $\mu \in [m_{P,S}/2, 2 m_{P,S}]$. We find that using  {\tt  NNPDF30\_nlo\_as\_0118\_lhcb} PDFs instead of the {\tt NNPDF30\_nlo\_as\_0118} set leads to a reduction of the total theoretical uncertainties by more than a factor of~2.  To assess the size of $P,S$ production from bottom-quark annihilation, we have employed the NLO corrections implemented in {\tt SusHi} \cite{Harlander:2012pb}. For resonance masses in the bottomonium region, we find that  $\sigma \hspace{0.25mm} \big ( b \bar b \to P,S \big ) \ll \sigma \hspace{0.25mm} \big ( gg \to P,S \big )$ and therefore neglect mediator production via bottom-quark annihilation in our numerical analysis. In view of this and given that known (approximate)  higher-order QCD corrections to $gg \to P,S$~\cite{Harlander:2002wh,Anastasiou:2002yz,Harlander:2002vv,Anastasiou:2002wq,Ravindran:2003um,Anastasiou:2014vaa} tend to increase the cross sections, we are confident that $\sigma \hspace{0.25mm} \big ( pp \to P \big ) > 6.3 \, {\rm nb}$ ($\sigma \hspace{0.25mm} \big ( pp \to S \big ) >  3.2 \, {\rm nb}$) in the mass region of interest. As will become clear in the next section, from these inclusive production rates  a non-negligible fraction of events falls into the LHCb acceptance, which covers pseudorapidities of $\eta \in [2.0, 4.5]$. 

From the above discussion and keeping in mind that  hadronic final states suffer from huge backgrounds and poor mass reconstruction, while measurements of diphoton final states are challenging at LHCb, it follows that looking for narrow resonances in dimuon decays seems to be the most promising search strategy at LHCb. In the following we will exploit this general idea by recasting the recent $\Upsilon (n)$ production measurements of LHCb~\cite{Aaij:2015awa} to derive bounds on the new-physics parameters entering~(\ref{eq:simplifiedmodel}). While this particular analysis is only sensitive to new dimuon resonances in the mass range $m_{P,S} \in [8.6, 12.4] \, {\rm GeV}$, extending the reach with future dedicated studies should be possible. Other previous studies of light spin-0 states using dimuon final states include~\cite{Lees:2012iw,Schmidt-Hoberg:2013hba,Dolan:2014ska,Aaij:2015tna,Dermisek:2009fd,Dermisek:2010mg,Bernon:2014nxa,Freytsis:2009ct,Batell:2009jf,Clarke:2013aya}.

\section{\boldmath Searching for peaks in the dimuon spectrum close to $\Upsilon (n)$}
\label{sec:searchstrategy}

Our discussion is based on the dimuon invariant mass spectrum at $\sqrt{s}=8 \, {\rm TeV}$ supplied as additional material and also presented in Fig.~1 (right) of the LHCb publication~\cite{Aaij:2015awa}. Using {\tt MadGraph5\_aMC@NLO}~\cite{Alwall:2014hca} generated  signal events for spin-0 resonances $\phi$, matched and showered with {\tt PYTHIA 6}~\cite{Sjostrand:2006za}, we first compute the~LHCb acceptance ($A$) based on the cuts $\eta \in [2.0, 4.5]$ and $p_T<30 \, {\rm GeV}$ that define the fiducial volume of the measurement. We obtain
\begin{equation} \label{eq:acceptance}
A = 0.23 \,,
\end{equation}
with negligible dependence on the $\phi$ mass within the experimental window $m_\phi \in [8.6,12.4]$~GeV and its parity. However, the available dimuon invariant mass spectrum data points correspond to a more restrictive kinematical region (i.e.~$\eta \in [3.0, 3.5]$ and  $p_T \in [3.0, 4.0] \, {\rm GeV}$) and the final acceptance $A_f$ does exhibit a mild dependence on~$m_\phi$. To cross-check  our results we have estimated the relative~$\Upsilon (n)$ acceptances $A_f/A$ by comparing the fitted~$\Upsilon (n)$ event yields in the dimuon spectrum to the corresponding measured fiducial cross sections.  We find that the relative acceptances of our generated signal and the LHCb $\Upsilon (n)$  production are similar when setting the~$\phi$ mass equal to that of $\Upsilon (n)$.  We also considered the LHCb mass resolution and its dependence on~$m_\phi$ by linearly interpolating/extrapolating the widths of the  fitted $\Upsilon (n)$  resonance shapes to higher and lower dimuon invariant masses.  These validations give us confidence that we understand the relative acceptances $A_f/A$ sufficiently well. 
 
\begin{figure}[!t]
\centering
\includegraphics[angle=0,width=7cm]{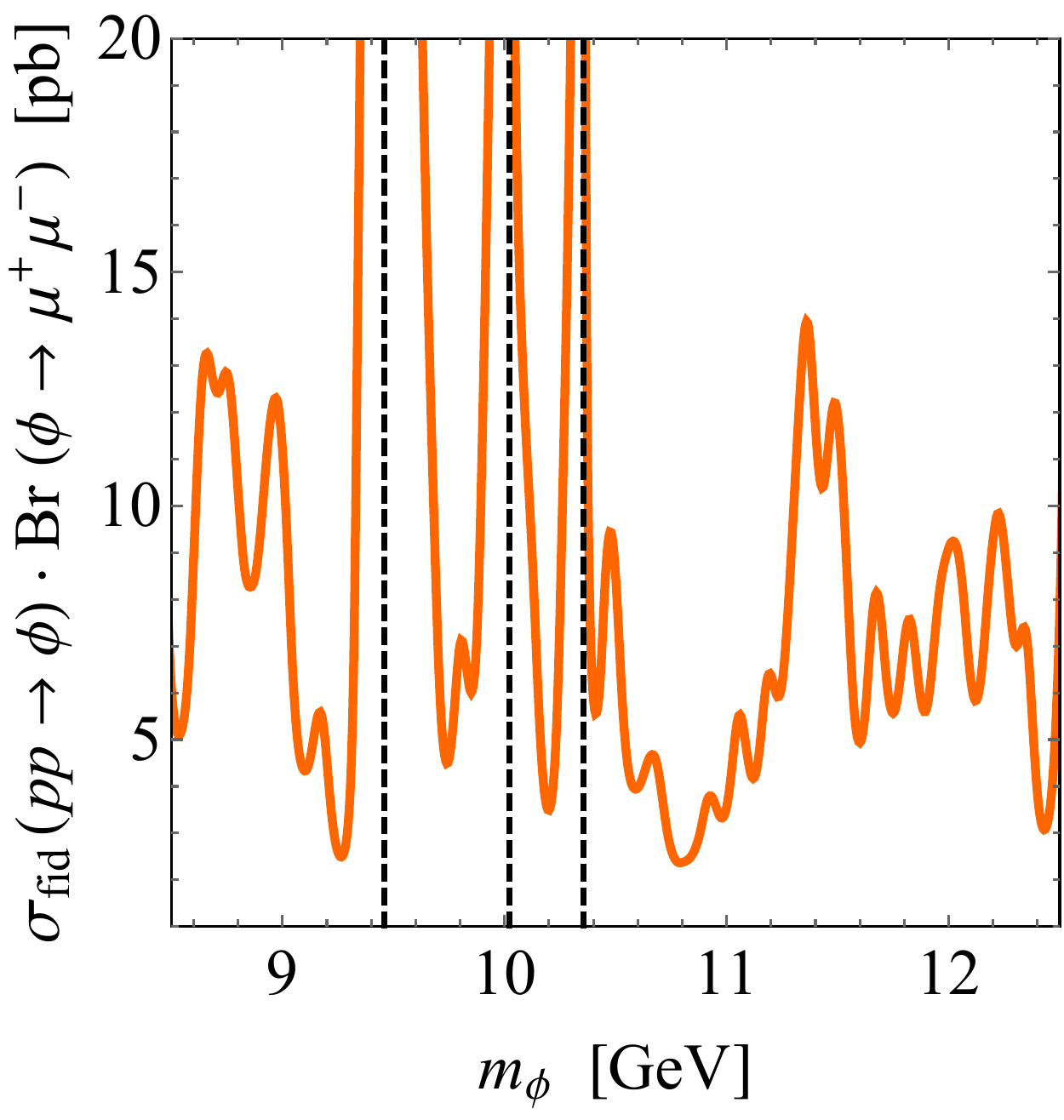}
\caption{\label{fig:xsecbound} $95\%$~CL bound on the fiducial production cross section times the dimuon branching ratio of an additional spin-0 resonance from a recast of the recent LHCb measurement of Upsilon production. The positions of the physical~$\Upsilon(n)$~($n=1,2,3$) masses are marked with black dashed lines. Consult the text for further details.}
\end{figure}

In order to constrain possible new-physics signals, we then refit the LHCb data while injecting an additional~$\phi$ resonance of a given mass, letting the normalisations of the existing~$\Upsilon (n)$ peaks vary freely, but keeping their positions fixed. Varying also the normalisation of the non-resonant background and performing a $\chi^2$ fit of the full spectrum, we extract the 95\%~CL bounds on the fiducial production cross section times dimuon branching ratio of an additional spin-0 resonance. Our final results are shown in Fig.~\ref{fig:xsecbound}. We observe that apart from the narrow regions around the~$\Upsilon(n)$ peaks of about $0.2 \, {\rm GeV}$, our method allows to constrain dimuon signal strengths $\sigma_{\rm fid} \hspace{0.25mm} \big (pp \to \phi  \big ) \cdot {\rm Br} \hspace{0.25mm} \big (\phi \to \mu^+ \mu^- \big )$ at the level of a few~pb. We believe that this is an interesting finding, in particular because the data set used to obtain these limits represents only around $3\%$ of all events recorded by LHCb at $8 \, {\rm TeV}$. Including the full data set available at LHC run I into the analysis is thus likely to significantly strengthen our proposal. In addition, future dedicated LHCb studies at run II that exploit our new search strategy should allow to extend the mass window to both lower and higher dimuon invariant masses.

\section{\boldmath Bounds on dimuon branching ratio of bottomonium states}
\label{sec:etabdimuonBr}

A simple but interesting application of the search strategy introduced in the last section is the derivation of upper limits on the dimuon branching ratios of the pseudoscalar $\eta_b (n)$ and scalar $\chi_b (n)$  bottomonium states. Let us illustrate this in the following. 

According to the non-relativistic QCD (NRQCD) factorisation approach~\cite{Bodwin:1994jh}, the prompt $\eta_b(n)$ production cross section in $pp$ collisions can be written as a convolution of the PDFs $f_{i,j/p} (x_{1,2})$ with the partonic cross sections~$\hat \sigma \hspace{0.25mm} \big (ij \to \eta_b(n) \big )$. The partonic cross section factorises  further  into perturbative coefficients that encode the production of a $b \bar b$ state and non-perturbative matrix elements  $\langle  {\cal O}^{\eta_b (n)} \rangle$ that describe the subsequent hadronisation of the $b \bar b$ pair into the observable $\eta_b (n)$ states. The matrix elements themselves can be expanded in powers of the relative velocity $v_b^2 \simeq 0.1$ of the $b$ quarks in the bottomonium system. 

\begin{table}[t!]
\centering
\begin{tabular}{c|@{\hspace{2mm}}c@{\hspace{2mm}}c@{\hspace{2mm}}|@{\hspace{2mm}}c@{\hspace{2mm}}c@{\hspace{2mm}}}
&  $m_{\eta_b(n)}$ & $\big |R_{\eta_b(n)} (0) \big |$ & $m_{\chi_b(n)}$ & $\big |R^\prime_{\chi_b (n)} (0) \big |$ \\[1mm]   
\hline  $n = 1$ & $9.4$ & $2.71 \pm 0.07$ &$9.86$ & $1.28 \pm 0.11$ \\
\hline  $n = 2$ & $10.0$ & $1.92 \pm 0.11$ & $10.23$ & $1.34 \pm 0.15$ \\
\hline  $n = 3$ & $10.3$ & $1.66 \pm 0.11$ &$10.51$ & $1.36 \pm 0.20$ \\
\hline  $n = 4$ & $10.6$ & $1.43 \pm 0.09$ & --- & --- \\
\hline  $n = 5$ & $10.85$ & $1.42 \pm 0.53$ & --- & --- \\
\hline  $n = 6$ & $11.0$ & $0.91 \pm 0.17$ & --- & --- 
\end{tabular}
\caption{Masses of the $\eta_b (n)$ $\big ( \chi_b (n) \big )$ states in units of ${\rm GeV}$ and corresponding values of the radial wave functions at the origin (their derivatives) in units of ${\rm GeV}^{3/2}$ (${\rm GeV}^{5/2}$).}
\label{tab:RRprime}
\end{table}

In fact, the production of $\eta_b(n)$ states is particularly simple, because  colour-octet Fock states  such as $b\bar b_8(^1S_0)$, $b\bar b_8(^3S_1)$ and  $b\bar b_8(^1P_1)$ are velocity suppressed compared to the colour-singlet $S$-wave contribution $b\bar b_1(^1S_0)$~\cite{Bodwin:1994jh}. The  $b\bar b_1(^1S_0)$ contribution hence fully dominates  $\eta_b (n)$ production \cite{Maltoni:2004hv}, and as a result the leading order (LO) gluon-gluon fusion cross sections are given by the following simple expression
\beq \label{eq:ggetabn}
\begin{split}
\hat \sigma \hspace{0.25mm} \big (g g \to \eta_b (n) \big ) & = \frac{\pi^3 \alpha_s^2}{36 m_b^3 \hat s} \, \delta \Big (1-\frac{4m_b^2}{\hat s} \Big ) \\[1mm] 
& \phantom{xx} \times \langle 0  | {\cal O}_1^{\eta_b (n)} (^1S_0) | 0 \rangle \,.
\end{split}
\eeq
Here the strong coupling constant $\alpha_s$  is understood to be evaluated at a scale $\mu_R \simeq 2 m_b$ with $m_b \simeq 4.75 \, {\rm GeV}$ the bottom pole mass, while $\hat s$ denotes the partonic centre of mass energy. The colour-singlet vacuum matrix elements are related to the $S$-wave radial wave functions at the origin via
\beq \label{eq:O11S0}
\langle 0  | {\cal O}_1^{\eta_b (n)}  (^1S_0) | 0 \rangle  = \frac{3}{2 \pi} \big | R_{\eta_b(n)} (0) \big |^2 \,,
\eeq
and the latter quantities can be extracted from the $\Upsilon(n)$ leptonic decay widths (see for instance \cite{Braaten:2000cm}) that are measured accurately \cite{Agashe:2014kda}. We collect 
the $\big |R_{\eta_b(n)} (0) \big |$ values that are used in our numerical analysis in Table~\ref{tab:RRprime}. Other determinations coming for example from potential models~\cite{Eichten:1995ch} agree  with our extractions within uncertainties. 

NLO QCD corrections to $\eta_b (n)$ hadroproduction have been calculated in \cite{Petrelli:1997ge} and include virtual corrections to the $gg \to b\bar b_1(^1S_0)$ channel as well as real processes, such as $gg \to  b\bar b_1(^1S_0) g$ or $gq \to  b\bar b_1(^1S_0) q$. Employing again {\tt NNPDF30\_nlo\_as\_0118\_lhcb} PDFs, we obtain the following ${\cal O} (\alpha_s^3)$ prediction for the inclusive  $\eta_b (n)$ production  cross sections 
\beq \label{eq:xsecppetab(n)}
\sigma \hspace{0.25mm} \big (p p \to \eta_b (n) \big) = \left ( 391^{+174}_{-68} \right ) \big | R_{\eta_b(n)} (0) \big |^2 \frac{{\rm n b}}{{\rm GeV}^3} \,,
\eeq
for $8 \, {\rm TeV}$ $pp$ collisions. The uncertainties quoted above include both the intrinsic PDF error as well as scale variations $\mu \in [m_b, 4 m_b]$ with  $\mu = \mu_R = \mu_F$. These two types of errors are are both asymmetric and similar in size. 

Equipped with the production rates (\ref{eq:xsecppetab(n)}) and having derived both the LHCb acceptance (\ref{eq:acceptance}) and the 95\%~CL limit on the dimuon signal strength in the last section, it is now straightforward to find upper bounds on the $\eta_b (n) \to \mu^+ \mu^-$ branching ratios. In the case of the lightest pseudoscalar bottomonium state, we find for instance 
\bea \label{eq:Bretab1dimuon}
 {\rm Br} \hspace{0.25mm} \big (  \eta_b (1) \to \mu^+ \mu^- \big ) < \frac{38.4 \, {\rm pb}}{\sigma \hspace{0.25mm} \big (p p \to \eta_b (1) \big)  \hspace{0.25mm} A} = 1.9 \cdot 10^{-4} \,. \hspace{3mm}
\eea
This limit has been obtained by applying the worst-case method \cite{Mistlberger:2012rs}, taking $882 \, {\rm nb}$ as the lowest possible $p p \to \eta_b (1)$ cross section. If  the uncertainty in~(\ref{eq:xsecppetab(n)}) and the error on $\big | R_{\eta_b(1)} (0) \big |$ as reported in Table~\ref{tab:RRprime} are combined in quadrature this lower bound  can be interpreted as a theoretical 95\%~CL limit. Notice that even with such a conservative treatment of uncertainties, our limit (\ref{eq:Bretab1dimuon}) is stronger by almost a factor of 50 than the 90\%~CL bound of ${\rm Br} \hspace{0.25mm} \big (  \eta_b (1) \to \mu^+ \mu^- \big ) < 9 \cdot 10^{-3}$ quoted by the Particle Data Group~(PDG)~\cite{Agashe:2014kda}. For the higher  pseudoscalar bottomonium states $\eta_b (2)$, $\eta_b (3)$, $\eta_b (4)$ and $\eta_b (6)$ our approach leads to worst-case upper bounds on the dimuon branching ratios in the range of $[0.8, 7.3] \cdot 10^{-4}$. No limits on these branching fractions are provided by the PDG. In the case of  $\eta_b (5)$, on the other hand, the large uncertainty on  $\big | R_{\eta_b(5)} (0) \big |$ when extracted from $\Upsilon(5)$ data, does not allow to set a meaningful bound. 

For the further discussion, we also need predictions for the production cross sections of the bottomonium scalar states $\chi_b(n)$. Unlike in the case of $\eta_b (n)$, two Fock states, namely $b\bar b_1(^3P_0)$ and $b\bar b_8(^3S_1)$, contribute to $\chi_b(n)$ hadroproduction at the same order in the velocity expansion of NRQCD~\cite{Bodwin:1994jh}.  The colour-octet configuration can however be produced at ${\cal O} (\alpha_s^2)$ only via $q \bar q \to b\bar b_8(^3S_1)$, while in the case of the colour singlet the process $g g \to b\bar b_1(^3P_0)$ is possible. Given the high gluon luminosities at the LHC, LO $\chi_b (n)$ production is hence described to very high accuracy  by 
\beq \label{eq:ggchibn}
\begin{split}
\hat \sigma \hspace{0.25mm} \big (g g \to \chi_b (n) \big ) & = \frac{\pi^3 \alpha_s^2}{4 m_b^5 \hat s} \, \delta \Big (1-\frac{4m_b^2}{\hat s} \Big ) \\[1mm] 
& \phantom{xx} \times \langle 0  | {\cal O}_1^{\chi_b (n)} (^3P_0) | 0 \rangle \,,
\end{split}
\eeq
with 
\beq \label{eq:O13P0}
\langle 0  | {\cal O}_1^{\chi_b (n)}  (^3P_0) | 0 \rangle  = \frac{9}{2 \pi} \big | R^\prime_{\chi_b(n)} (0) \big |^2 \,.
\eeq
The derivatives of the $P$-wave radial wave functions at the origin $ | R^\prime_{\chi_b(n)} (0) \big |$ cannot be extracted from experiment and one thus has to rely on theory to obtain their values. As estimates of the derivatives of the radial wave functions, we take the mean values from the four potential-model calculations presented in \cite{Eichten:1995ch}. The numerical values that we employ in our work are tabulated in  Table~\ref{tab:RRprime}. The uncertainties given in this table are the standard deviations that derive from the results of the four different potential-model computations. 

Like in the case of the $\eta_b(n)$ states, ${\cal O} (\alpha_s^3)$ corrections to prompt $pp \to \chi_b(n)$ production are important and have been calculated~\cite{Petrelli:1997ge}. Adopting the same methodology that led to~(\ref{eq:xsecppetab(n)}), we obtain at  $8 \, {\rm TeV}$ the NLO result 
\beq \label{eq:xsecppchib(n)}
\sigma \hspace{0.25mm} \big (p p \to \chi_b (n) \big) = \left ( 504^{+417}_{-169} \right ) \big | R^\prime_{\chi_b(n)} (0) \big |^2 \frac{{\rm n b}}{{\rm GeV}^5} \,.
\eeq
Notice that, as a result of the sizeable $gq \to  b\bar b_1(^3P_0) q$ contribution which first contributes to $\chi_b(n)$ hadroproduction at  ${\cal O} (\alpha_s^3)$, the uncertainties plaguing (\ref{eq:xsecppchib(n)}) are more than twice as large as those entering (\ref{eq:xsecppetab(n)}). The above prompt  production cross sections can again be translated into lower limits on the $\chi_b (n)$ dimuon branching ratios.  At the 95\%~CL, we obtain  ${\rm Br} \hspace{0.25mm} \big ( \chi_b (n) \to \mu^+ \mu^- \big)$ values in the range of $[1.3, 4.0] \cdot 10^{-5}$.

The bounds on the dimuon branching ratios that we have derived  should be compared to the corresponding SM expectations. Using  the formulas given in~App.~\ref{app:widths}, we obtain ${\rm Br} \hspace{0.25mm} \big (  \eta_b (1) \to \mu^+ \mu^- \big ) \simeq {\rm Br} \hspace{0.25mm} \big (  \eta_b (1) \to Z^\ast \to  \mu^+ \mu^- \big ) \simeq 2 \cdot 10^{-10}$ and ${\rm Br} \hspace{0.25mm} \big (  \chi_b (1) \to \mu^+ \mu^- \big ) \simeq {\rm Br} \hspace{0.25mm} \big (  \chi_b (1) \to \gamma^\ast \gamma^\ast \to \mu^+ \mu^- \big ) \simeq 7 \cdot 10^{-11}$. Similar results also hold for all other spin-0 bottomonium states with masses below the open bottom threshold. Our limits are thus around six orders of magnitude above the SM expectations.

\section{Bounds on new spin-0 dimuon resonances}
\label{sec:spin0bounds}

At this point, we have collected all ingredients necessary to interpret the bounds derived in Sec.~\ref{sec:searchstrategy} in terms of new spin-0 states described by \eqref{eq:simplifiedmodel} or more specific models like the THDMII. In doing so we need to consider non-perturbative effects due to the presence of  bottomonium resonances and the $b\bar b$ threshold. In particular, close to the $b\bar b$ threshold a perturbative description of the production and the decay of the new resonances breaks down. In this region we can however approximate the $b\bar b$ contributions to the $P,S$ widths by a sum over exclusive states interpolated to the continuum sufficiently above threshold~\cite{Drees:1989du,Baumgart:2012pj}. Like these analyses, we also assume that the dominant contributions to production and the total width arise from the mixing of the new spin-0 mediators with bottomonium states. In particular, $P$ will mix with the  six $\eta_b (n)$ states, while $S$ will mix with the three $\chi_b(n)$ resonances. Such mixings can effectively be described through off-diagonal contributions $\delta m^2_{P\eta_b(n)}$ to the pseudoscalar mass matrix squared
\begin{widetext}
\beq \label{eq:masssquared}
M_{P\eta_b}^2 = \left( 
\begin{array}{cccc} 
m_P^2 - i m_P \Gamma_P & \delta m^2_{P\eta_b(1)} & \ldots  &  \delta m^2_{P\eta_b(6)} \\
 \delta m^2_{P\eta_b(1)} &  m_{\eta_b(1)}^2 - i m_{\eta_b(1)} \Gamma_{\eta_b(1)} & \ldots & 0 \\
 \vdots & 0 & \ddots & 0 \\
\delta m^2_{P\eta_b(6)} & 0 & 0 & m_{\eta_b(6)}^2 - i m_{\eta_b(6)} \Gamma_{\eta_b(6)} 
  \end{array}  \right) \,,
\eeq
\end{widetext}
and its analogue $M^2_{S\chi_b}$ in the scalar case. The masses  of the $\eta_b (n)$ $\big ($$\chi_b (n)$$\big )$ states are denoted by $m_{\eta_b(n)}$ $\big ($$m_{\chi_b(n)}$$\big )$ and their numerical values are collected in Table~\ref{tab:RRprime}. The total decay widths  $\Gamma_{\eta_b(n)}$ $\big ($$\Gamma_{\chi_b(n)}$$\big )$ of the unmixed  pseudoscalar (scalar) bottomonium states are calculated using  the formalism employed in~\cite{Drees:1989du,Baumgart:2012pj}.  The relevant expressions are given in~App.~\ref{app:widths}. The formulas needed to predict the total decay widths $\Gamma_{P,S}$ of the new spin-0 resonances can also be found there. 

The off-diagonal entries  appearing in (\ref{eq:masssquared}) can be computed using NRQCD~\cite{Bodwin:1994jh}. To zeroths order in $\alpha_s$ and $v_b$, one recovers the non-relativistic potential model results~(see for instance~\cite{Drees:1989du})
\beq \label{eq:offdiagonal}
\begin{split}
\delta m^2_{P\eta_b(n)}  &= \kappa_P^b \hspace{0.25mm} \sqrt{\frac{3 }{4\pi v^2}  \hspace{0.25mm} m_{\eta_b(n)}^3 }\hspace{0.25mm}   \big |R_{\eta_b(n)}(0) \big  |\,, \\[2mm]
\delta m^2_{S\chi_b(n)}  &= \kappa_S^b  \hspace{0.25mm} \sqrt{\frac{27}{\pi v^2}  \hspace{0.25mm} m_{\chi_b(n)} } \hspace{0.25mm}  \big |R^\prime_{\chi_b(n)}(0) \big |\,.
\end{split}
\eeq
We identify the physical $P,S$ states with the ones containing the largest $P,S$ admixture. Their total decay widths can be read off directly from the imaginary parts of the  mass matrix squared (\ref{eq:masssquared}) after diagonalisation. The small mass-shift effects~\cite{Drees:1989du} needed to avoid   level crossing when the $P,S$ mass (before mixing) is close to any of the considered bottomonium states are, on the other hand, neglected in our analysis. 

\begin{figure}[t!]
\centering
\vskip0.3cm
\includegraphics[angle=0,width=7.5cm]{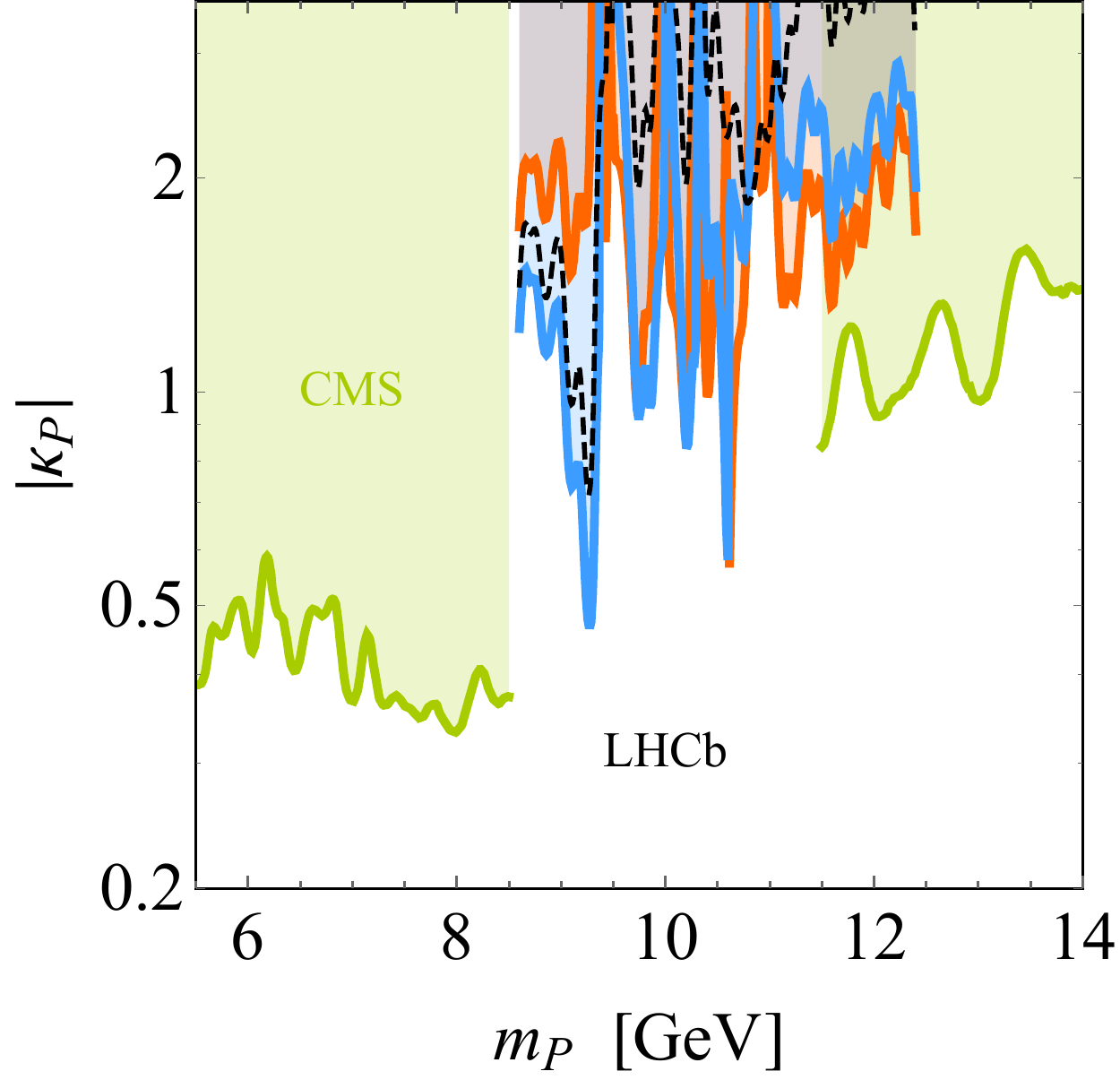}
\vskip0.5cm
\includegraphics[angle=0,width=7.5cm]{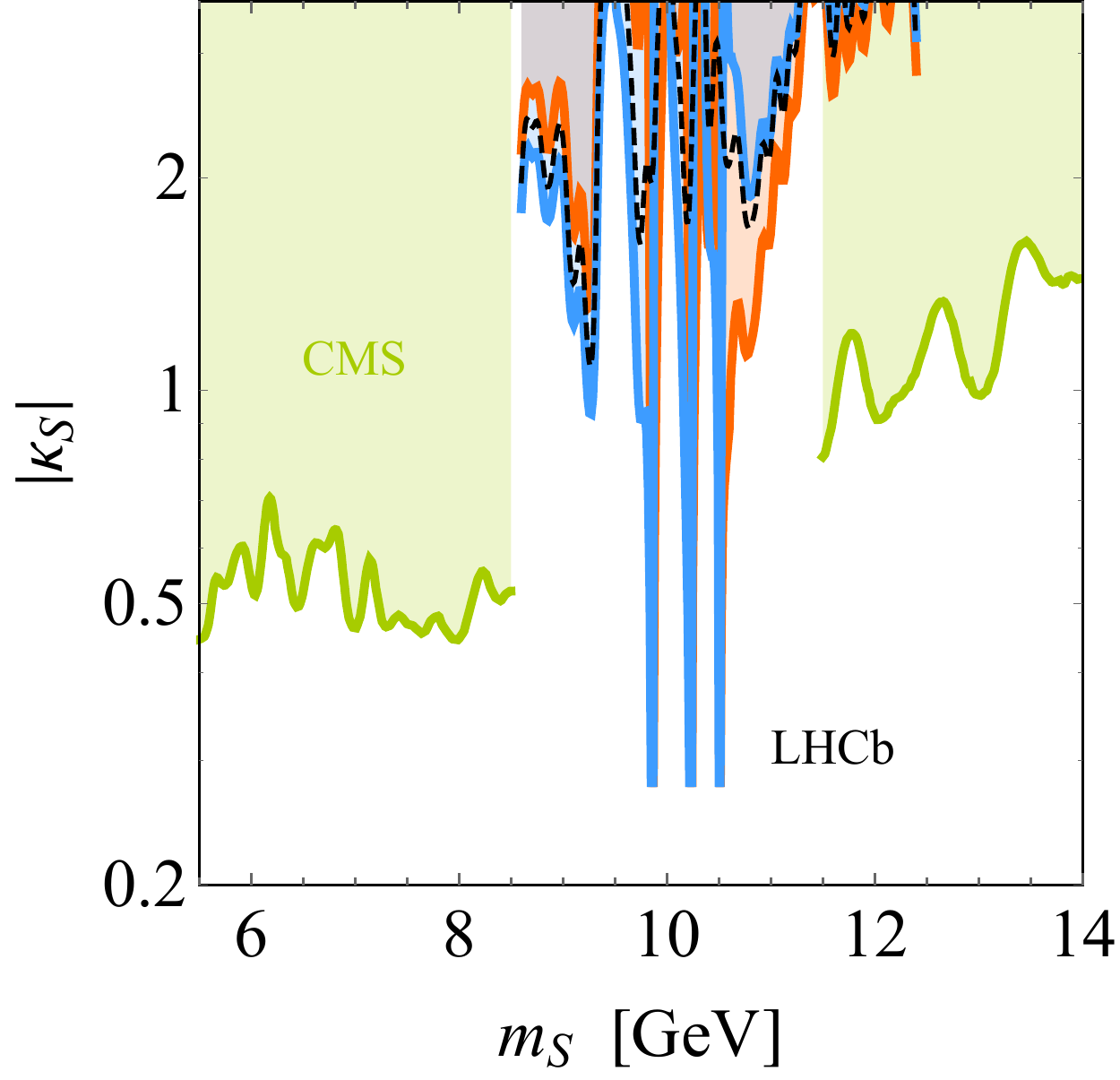}
\caption{\label{fig:bounds} 95\%~CL limits on the universal coupling strengths~$|\kappa_{P}|$~(upper panel) and $|\kappa_{S}|$ (lower panel). The red, blue and black curves are obtained from our recast of the measurement of $\Upsilon (n)$ production at LHCb, while the green curves stem from a resonance search in the dimuon channel performed by~CMS. The shaded regions correspond to disfavoured parameter space. See text for additional explanations. }
\end{figure}

The mass mixing has the most significant effect on the couplings of the spin-0 mediators to gluons modifying both the production and the total decay width of the physical (i.e.~mixed) $P,S$ states. The associated interference in production is affected by the strong phase present in the $gg \to P,S$ amplitudes due to intermediate on-shell charm and bottom quarks. Numerically more important than the strong phases are the signs of the couplings $\kappa^{c,b,t}_{P}$~$\big($$\kappa^{c,b,t}_{S}$$\big)$ relative to $R_{\eta_b(n)}$~$\big($$R^\prime_{\chi_b(n)}$$\big )$, since these signs determine the interference pattern between the new resonances $P,S$ and the QCD spin-0 bound states.  For instance, close to the bottomonium resonances, $P,S$ production can be significantly enhanced as an effect of mixing. At the same time however, the decay mode $P,S \to gg$ then tends to dominate the total width of the new state, which leads to a further suppression of its dimuon branching ratio. In an accurate calculation, both effects need to be taken into account.  Following the discussion in Sec.~\ref{sec:etabdimuonBr}, bottomonium contributions to the dimuon partial widths of the  physical $P,S$ states are numerically insignificant and can be ignored. 

Before presenting our numerical results, we finally note that since the mixing contributions (\ref{eq:offdiagonal}) are proportional  to  $\kappa_{P,S}^b$, their impact  on the phenomenology of the $P,S$ states diminishes (and the effects become more localised to the bottomonium thresholds) as the bounds on the couplings in the simplified model (\ref{eq:simplifiedmodel}) become stronger.  Conversely, it turns out that for $\kappa_{P,S}^b\gtrsim \sqrt{4\pi}$ the large mixings and resulting total $P,S$ decay widths start making the identification of physical $P,S$ resonances ambiguous as simultaneous mixing with several bottomonium states becomes important. To avoid this issue, we restrict our analysis to the parameter region $\kappa_{P,S}^b \lesssim \sqrt{4\pi}$.  

In Fig.~\ref{fig:bounds} we present the limits on the magnitudes of the universal couplings~$\kappa_{P,S} = \kappa_{P,S}^f$ by  employing the 95\%~CL bounds on the dimuon LHCb signal strength derived in~Sec.~\ref{sec:searchstrategy}. The shown exclusions are based on the conservative lower bounds of our cross section calculations shown in Fig.~\ref{fig:xsecs} and given in (\ref{eq:xsecppetab(n)}) and (\ref{eq:xsecppchib(n)}). For comparison, predictions with (red and blue curves) and without (black dashed curves) mixing effects are presented. In the case of mixing, we consider both relative signs of the couplings $\kappa_{P,S}$ to illustrate the associated model dependence. The red (blue)  curves correspond to the case where the sign of $\kappa_{P,S}$ is such that the interference in the physical $P,S \to gg$ decay is destructive (constructive) for $m_{P,S} < m_{\eta_b (n), \hspace{0.25mm} \chi_b (n)}$. From the panels it is evident that while mixing effects play a particular important role in the pseudoscalar case, due to the large number of QCD resonances and the more pronounced mass mixing $\delta m^2_{P\eta_b(n)} $, the obtained bounds are also changed  in the scalar case as a result of $\delta m^2_{S\chi_b(n)}  \neq 0$. In addition, effects associated to $P \to B^\ast \bar B$ $\big ($see (\ref{eq:BastB})$\big)$ are phenomenologically important, since they strengthen the limits on $|\kappa_{P}|$ visibly for $m_P \in [11 , 12.4] \, {\rm GeV}$. In the scalar case such effects are instead of minor importance. One finally notices that our proposal allows to set the first relevant limits of ${\cal O} (1)$ on the coupling strengths $|\kappa_{P,S}|$ for $m_{P,S} \in [8.6, 11.5] \, {\rm GeV}$. This mass range has so far not been covered  by other analyses such as the CMS dimuon search~\cite{Chatrchyan:2012am}, which provides the strongest constraints on  $|\kappa_{P,S}|$ for $m_{P,S} \in [5.5, 8.6] \, {\rm GeV}$ and $m_{P,S} \in [11.5, 14] \, {\rm GeV}$ (green curves). The recent LHCb precision measurement of $\Upsilon (n)$ production thus enables one to close a gap in parameter space. 

\section{Discussion and outlook}
\label{sec:conclusions}

\begin{figure}[t!]
\centering
\includegraphics[angle=0,width=7.5cm]{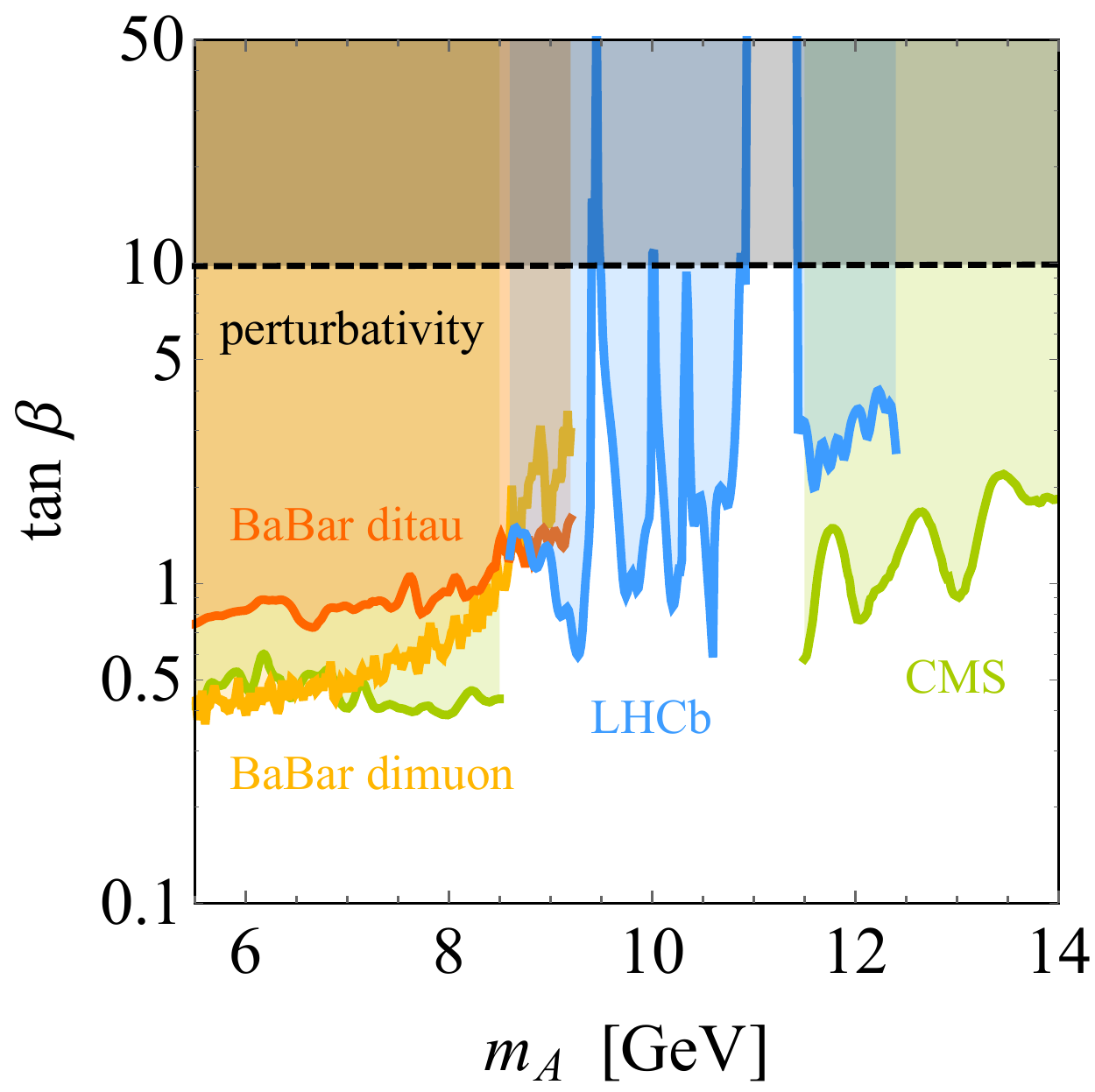}
\caption{\label{fig:boundTB} 95\%~CL bound on $\tan \beta$ in the THDMII scenario. The blue curve follows from $\Upsilon (n)$ production by LHCb, the green curve arises from  a  CMS dimuon resonance search, whereas the yellow and orange curve derives from the  BaBar 90\%~CL limit on radiative $\Upsilon (1)$ decays in the dimuon and ditau channel, respectively.  The bound on $\tan \beta$ arising from perturbativity is also shown~(black dashed line). All shaded regions correspond to excluded parameter space. For further details see main text. }
\end{figure}

The generic bounds on new light spin-0 states presented in the previous section can be easily interpreted within  ultraviolet complete new-physics models such as THDM scenarios or the next-to-minimal supersymmetric SM. As an example, we show  in Fig.~\ref{fig:boundTB} the  limits on $\tan \beta$ in the  decoupling limit of the THDMII for pseudoscalar masses $m_A$ close to $10 \, {\rm GeV}$ following from our recast~(blue curve),  the CMS dimuon search~\cite{Chatrchyan:2012am}~(green curve) and the  BaBar limit on radiative $\Upsilon (1) \to \gamma A$ decays with $A \to \mu^+ \mu^-$~\cite{Lees:2012iw}~(yellow curve) or $A \to \tau^+ \tau^-$~\cite{Lees:2012te}~(orange curve). For comparison, we also indicate~(black dashed line) the parameter space consistent with perturbativity of the scalar potential (cf.~\cite{Bernon:2014nxa} for a recent discussion). The shown LHCb bound has been obtained by incorporating the full mixing effects described in Sec.~\ref{sec:spin0bounds} and taking the interference pattern in the $A \to gg$ decay to be constructive for $m_A < m_{\eta_b (n)}$~\cite{Drees:1989du}. From the figure, one observes that the existing analyses of dimuon and ditau final states  provide stringent constraints on the THDMII in almost the entire low-$m_A$ mass range, with our recast of the recent LHCb~$\Upsilon(n)$ production measurement furnishing the dominant restriction for $m_A \in [8.6, 11] \, {\rm GeV}$. Only the masses $m_A \in [11, 11.5] \, {\rm GeV}$ remain unexplored, since mixing effects turn out to be particularly important in this region. We finally recall that the LHCb data used in our fit correspond to only  $3\%$ of all recorded dimuon events. Consequently, a dedicated LHCb analysis of the full run I data set is expected to improve  the limits derived here considerably, possibly allowing to surpass  the existing CMS constraints for $m_A > 11.5 \, {\rm GeV}$.

While we have focused  in our work on the  experimentally cleanest  signature, namely dimuons, light spin-0 resonances could also be searched for in other final states like $\tau^+\tau^-$, $c\bar c$ or  $b\bar b$. All these modes do benefit from larger branching ratios (see Fig.~\ref{fig:BrGa}), but compared to $\mu^+ \mu^-$ do suffer from much more challenging reconstruction and considerably larger backgrounds.  The $P,S \to \tau^+ \tau^-$ decay in particular seems less promising, because the visible charged particles in the decay are a bad proxy for the total momentum of the taus, with most of the energy  typically being carried away by the neutrinos. The resulting invariant mass distribution thus has no pronounced peak for $m_{P,S} \lesssim 15 \, {\rm GeV}$. In this respect, LHCb searches for resonances in the exclusive invariant mass spectra of heavy flavoured hadrons, such as $D^+ D^-$ or $B^+ B^-$ may have more potential. A dedicated study of the corresponding phenomenology, while  beyond the scope of this work, might hence turn out to be a fruitful exercise. 

\begin{acknowledgments}

We thank the authors of \cite{Gauld:2015yia} for a {\tt LHAPDF6} version~\cite{Buckley:2014ana} of the {\tt NNPDF30\_nlo\_as\_0118\_lhcb} set and Matteo Cacciari for a {\tt Fortran} implementation of the results presented in~\cite{Petrelli:1997ge}.  A big thank you to Tobias~Golling, Philip~Ilten, Andrey~Katz and Michael~Spira for insightful discussions and Daniel~Maitre for prompt feedback concerning~{\tt HPL}~\cite{Maitre:2005uu, Maitre:2007kp}. UH acknowledges the hospitality and support of the CERN theory division. The work of JFK was supported in part  by the Slovenian Research Agency. 

\end{acknowledgments}

\begin{appendix}

\begin{widetext}

\section{Decay width formulas}
\label{app:widths}

In our numerical evaluation of the total decay  widths of the new spin-0  states, we evaluate the quark masses appearing in the interactions (\ref{eq:simplifiedmodel}) in the  $\overline{\rm MS}$ scheme at the renormalisation scale $\mu_R = m_{P,S}$. Masses related to kinematic (i.e.~phase-space) effects in $\Gamma_{P,S}$ are instead evaluated using the pole scheme. In the pseudoscalar case, the partial decay widths are given by (see for instance \cite{Djouadi:1996yq,Djouadi:2005gi,Djouadi:2005gj,Haisch:2012kf})
\begin{align}
\Gamma\hspace{0.25mm} (P\to q\bar q) & = \frac{\big |\kappa_P^q \big |^2 d_q  \hspace{0.25mm} m_q^2 \hspace{0.25mm}  m_P}{8\pi v^2} \hspace{0.5mm}  \beta_{q/P} \left  (1+\Delta_P^q \right )\,, \\[1mm]
\Gamma\hspace{0.25mm} (P\to \ell^+ \ell^-) & = \frac{\big |\kappa_P^\ell \big |^2 d_\ell   \hspace{0.35mm}  m_\ell^2 \hspace{0.25mm}  m_P}{8\pi v^2} \hspace{0.5mm}   \beta_{\ell/P}  \,, \\[1mm]
\Gamma\hspace{0.25mm} (P\to gg) & = \frac{\alpha_s^2 \hspace{0.25mm} m_P^3}{32\pi^3 v^2} \left| \sum_{q} \kappa_P^q  \hspace{0.5mm}  \big( \mathcal P (\tau_P^q)  +   \Delta_P^g \big ) \right|^2  \,, \\[1mm]
\Gamma\hspace{0.25mm} (P\to \gamma\gamma) & = \frac{\alpha^2 \hspace{0.25mm} m_P^3}{64\pi^3 v^2} \left| \sum_f \kappa_P^f d_f Q_f^2 \hspace{0.5mm}  \big ( \mathcal P (\tau_P^f)+  \Delta_P^\gamma \big ) \right|^2 \,, \\[1mm]
\Gamma\hspace{0.25mm} (P\to \gamma Z) & = \frac{\alpha^2 m_P^3}{32 \pi^3 v^2}  \left  (1- \frac{m_Z^2}{m_P^2} \right )^3 \left| \sum_f \kappa_P^f d_f Q_f \, \frac{I_f - 2 Q_f s_w^2}{s_w c_w}\,  {\cal I} (\tau_Z^f, \tau_P^f) \right|^2\,.
\end{align}
Here $\tau_i^{j} = (2 m_{j} /m_i)^2$, $\beta_{i/j} = \sqrt{1 - \tau_j^{i}}$, $d_q =3$ ($d_\ell =1$) is the dimension of the fermionic colour representation for quarks (leptons), $I_f$ is the third component of the weak isospin of the relevant fermion, while $Q_f$ denotes its electric charge. Finally,  $s_w = \sin \theta_w$ and $c_w = \cos \theta_w$ are the sine and cosine of the weak mixing angle, respectively. 

For scalar particles, we employ 
\begin{align}
\Gamma\hspace{0.25mm} (S\to q\bar q) & = \frac{\big |\kappa_S^q \big |^2 d_q  \hspace{0.25mm} m_q^2 \hspace{0.25mm}  m_S}{8\pi v^2} \hspace{0.5mm}  \beta_{q/S}^{3} \left (1+\Delta_S^q  \right )\,, \\[1mm]
\Gamma\hspace{0.25mm} (S\to \ell^+ \ell^-) & = \frac{\big |\kappa_S^\ell \big |^2 d_\ell  \hspace{0.35mm} m_\ell^2 \hspace{0.25mm}  m_S}{8\pi v^2} \hspace{0.5mm}  \beta_{\ell/S}^{3} \,, \\[1mm]
\Gamma\hspace{0.25mm} (S\to gg) & = \frac{\alpha_s^2 \hspace{0.25mm} m_S^3}{32\pi^3 v^2} \left| \sum_{q} \kappa_S^q \hspace{0.5mm} \big (  \mathcal S (\tau_S^q)  +    \Delta_S^g \big ) \right|^2 \,, \\[1mm]
\Gamma\hspace{0.25mm} (S\to \gamma\gamma) & = \frac{\alpha^2 \hspace{0.25mm} m_S^3}{64\pi^3 v^2} \left| \sum_f \kappa_S^f d_f Q_f^2 \hspace{0.5mm} \big (  \mathcal S(\tau_S^f)  + \Delta_S^\gamma \big ) \right|^2  \,, \\[1mm]
\Gamma\hspace{0.25mm} (S\to \gamma Z) & = \frac{\alpha^2 m_S^3 }{32 \pi^3 v^2} \left (1-\frac{m_Z^2}{m_S^2} \right )^3 \left| \sum_f \kappa_P^f d_f Q_f \, \frac{I_f - 2 Q_f s_w^2}{s_w c_w} \left [ {\cal J} (\tau_Z^f, \tau_S^f)- {\cal I} (\tau_Z^f, \tau_S^f) \right ] \right|^2\,.
\end{align}

The relevant loop functions take the form 
\begin{align}
\mathcal P (\tau) &= \tau \arctan^2 \left( \frac{1}{\sqrt{\tau-1}} \right) \,, \\[1mm]
\mathcal S(\tau) &= \tau + \left (1-\tau \right ) {\cal P}(\tau)  \,, \\[2mm]
\mathcal T(\tau) &= \sqrt{\frac{\tau-1}{\tau} \, {\cal P}(\tau)} \,, \\[1mm]
\mathcal I (\tau_i, \tau_j) &= -\frac{\tau_i \tau_j}{2(\tau_i-\tau_j)} \left [ \frac{{\mathcal P} (\tau_i)}{\tau_i} - \frac{{\mathcal P} (\tau_j)}{\tau_j} \right]\, , \\[1mm]
\mathcal J (\tau_i, \tau_j) &= \frac{\tau_i \tau_j}{2(\tau_i-\tau_j)}  + \frac{\tau_i^2\tau_j^2}{2(\tau_i-\tau_j)^2} \left [ \frac{{\mathcal P} (\tau_i)}{\tau_i} - \frac{{\mathcal P} (\tau_j)}{\tau_j} \right ] +  \frac{\tau_i\tau_j^2}{(\tau_i-\tau_j)^2} \, \big [ \mathcal T (\tau_i) - \mathcal T (\tau_j) \big ] \,. 
\end{align}

The perturbative QCD corrections to the partial widths into quark pairs that we include in our analysis are (cf.~\cite{Drees:1989du})
\begin{align}
\Delta_P^q & =  \frac{4 \alpha_s}{3 \pi} \left ( \frac{{\cal Q} (\beta_{q/P})}{\beta_{q/P}}  - \frac{19+2\beta_{q/P}^2+3 \beta_{q/P}^4 }{16 \beta_{q/P}} \ln x_{\beta_{q/P}} + \frac{21  - 3 \beta_{q/P}^2}{8} \right ) \,,\\[2mm]
\Delta_S^q & = \frac{4 \alpha_s}{3 \pi} \left ( \frac{{\cal Q} (\beta_{q/S})}{\beta_{q/S}}  - \frac{3+34\beta_{q/S}^2-13 \beta_{q/S}^4 }{16 \beta_{q/S}^3} \ln x_{\beta_{q/S}} - \frac{3  - 21 \beta_{q/S}^2}{8 \beta_{q/S}^2} \right )  \,,
\end{align}
where we have introduced the abbreviation $x_{\beta_{i/j}} = (1 - \beta_{i/j})/(1 + \beta_{i/j})$ and the function ${\cal Q} (\beta)$ takes the form 
\beq
{\cal Q} (\beta) = \big (1+\beta^2 \big )  \left ( 4 \hspace{0.25mm} {\rm Li}_2 (x_\beta) + 2  \hspace{0.25mm}  {\rm Li}_2 (-x_\beta) + 3 \ln x_\beta \ln \frac{2}{1+\beta} + 2 \ln x_\beta \ln \beta \right ) - 3 \beta \ln \frac{4 \beta^{4/3}}{1-\beta^2} \,,
\eeq
with ${\rm Li}_2 (z)$ the usual dilogarithm.

The QCD corrections to the digluon partial widths can be written as ($\phi = P,S$)
\begin{equation} \label{eq:digluon}
\Delta_{\phi}^g  =  \frac{\alpha_s}{\pi}  \left(  {\cal G}_\phi (y_{\beta_{q/\phi}})+  {\cal M}_\phi (\tau_\phi^q) \hspace{0.25mm} \ln \frac{\mu_q^2}{m_q^2}  \right)\,,
\end{equation}
where $y_{\beta_{i/j}}= -x_{\beta_{i/j}}$ and we set $\mu_q = m_\phi/2$ in our analysis to  reproduce the position of the $q \bar q$ decay threshold correctly. The loop functions appearing above take the form \cite{Harlander:2005rq}
\begin{align}
{\cal G}_P (y) & = \frac{y}{ {\left(1-y\right)}^2}\biggl[ 48\HPL(1,0,-1,0;y)+ 4\ln(1-y)\ln^3 y- 24\hspace{0.25mm} \zeta_2 \Li_2( y)- 24 \hspace{0.25mm} \zeta_2 \ln(1- y)\ln y - 72 \hspace{0.25mm}\zeta_3 \ln(1- y) \nonumber \\[1mm]
& \hspace{2cm} - \frac{220}{3}\Li_3( y) - \frac{128}{3}\Li_3(- y) + 68\Li_2( y)\ln y + \frac{64}{3}\Li_2(- y)\ln y + \frac{94}{3}\ln(1- y)\ln^2 y \nonumber \\[1mm] 
& \hspace{2cm}  - \frac{16}{3}\hspace{0.25mm}\zeta_2\ln y + \frac{124}{3}\hspace{0.25mm}\zeta_3 + 3\ln^2 y \biggr]   - \frac{ 24  y \left(5+7 { y}^2\right) }{{\left(1- y\right)}^3 \left(1+ y\right)}\Li_4( y)  -\frac{ 24 y \left(5+11 { y}^2\right)}{{\left(1- y\right)}^3 \left(1+ y\right)}\Li_4(-y) \nonumber \\[1mm]
\\[-5mm]
\nonumber 
\\[-5mm]
& \phantom{xx} +\frac{ 8 y \left(23+41 { y}^2\right) } {3{\left(1- y\right)}^3 \left(1+ y\right)} \biggl[ \Li_3( y) +\Li_3(- y) \biggr] \ln y -\frac{ 4 y \left(5+23 { y}^2\right) }{3{\left(1- y\right)}^3 \left(1+ y\right)}\Li_2( y)\ln^2 y  \nonumber \\[1mm] 
& \phantom{xx} - \frac{ 32 y \left(1+{ y}^2\right) }{3{\left(1- y\right)}^3 \left(1+ y\right)}\Li_2(- y)\ln^2 y  +\frac{  y \left(5-13 { y}^2\right) }{36 {\left(1- y\right)}^3 \left(1+ y\right)}\ln^4 y +\frac{ 2 y \left(1-17 { y}^2\right) }{3{\left(1- y\right)}^3 \left(1+ y\right)}\hspace{0,25mm} \zeta_2\ln^2 y \nonumber \\[1mm] 
& \phantom{xx} +\frac{ 4 y \left(11-43 { y}^2\right) }{3{\left(1- y\right)}^3 \left(1+ y\right)} \hspace{0.25mm} \zeta_3\ln y +\frac{ 24 y \left(1-3 { y}^2\right) }{{\left(1- y\right)}^3 \left(1+ y\right)} \hspace{0.25mm} \zeta_4 +\frac{ 2 y \left(2+11 y\right) }{ 3{\left(1- y\right)}^3}\ln^3 y \,, \nonumber \\[2mm]
{\cal G}_S (y) & = \frac{  y {\left(1+ y\right)}^2 }{{\left(1- y\right)}^4}  \biggl[ 72	\HPL(1,0,-1,0; y) + 6\ln(1- y)\ln^3 y - 36\hspace{0.25mm} \zeta_2\Li_2( y) 	- 36 \hspace{0.25mm}\zeta_2\ln(1- y)\ln y    -108 \hspace{0.25mm} \zeta_3\ln(1- y) \nonumber \\[1mm] 
& \hspace{2cm} - 64\Li_3(- y) + 32\Li_2(- y)\ln y - 8\hspace{0.25mm}\zeta_2\ln y \biggr]  -\frac{ 36 y \left(5+5 y+11 { y}^2+11 { y}^3\right)}{{\left(1- y\right)}^5}\Li_4(- y)  \nonumber \\[1mm]
& \phantom{xx} -\frac{ 36 y \left(5+5 y+7 { y}^2+7 { y}^3\right)}{{\left(1- y\right)}^5}\Li_4( y) +\frac{ 4 y \left(1+ y\right) \left(23+41 { y}^2\right)}{{\left(1- y\right)}^5}  \biggl[ \Li_3( y) +\Li_3(- y) \biggr] \ln y \,, \nonumber \\[1mm]
& \phantom{xx} -\frac{ 16 y \left(1+ y+{ y}^2+{ y}^3\right)}{{\left(1- y\right)}^5}\Li_2(- y)\ln^2 y -\frac{ 2 y \left(5+5 y+23 { y}^2+23 { y}^3\right) }{{\left(1- y\right)}^5}\Li_2( y)\ln^2 y \nonumber \\[1mm]
& \phantom{xx}  +\frac{  y \left(5+5 y-13 { y}^2-13 { y}^3\right) }{24 {\left(1- y\right)}^5}\ln^4 y +\frac{  y \left(1+ y-17 { y}^2-17 { y}^3\right) }{{\left(1- y\right)}^5} \hspace{0.25mm} \zeta_2\ln^2 y \\[1mm]
& \phantom{xx}   +\frac{ 2 y \left(11+11 y-43{ y}^2-43{ y}^3\right) }{{\left(1- y\right)}^5}\hspace{0.25mm} \zeta_3\ln y  + \frac{ 36 y \left(1+ y-3 { y}^2-3 { y}^3\right) }{{\left(1-y \right)}^5} \hspace{0.25mm} \zeta_4 -\frac{ 2 y \left(55+82 y+55 { y}^2\right) }{{\left(1- y\right)}^4}\Li_3( y)  \nonumber \\[1mm]
& \phantom{xx}  +\frac{ 2 y \left(51+74 y+51{ y}^2\right) }{{\left(1- y\right)}^4}\Li_2( y)\ln y +\frac{  y \left(47+66 y+47 { y}^2\right) }{{\left(1- y\right)}^4}\ln(1- y)\ln^2 y \nonumber \\[1mm] 
& \phantom{xx} +\frac{  y \left(6+59 y+58 { y}^2+33 { y}^3\right) }{3 {\left(1- y\right)}^5}\ln^3 y +\frac{ 2 y \left(31+34 y+31 { y}^2\right)}{{\left(1- y\right)}^4}\hspace{0.25mm} \zeta_3+\frac{ 3 y \left(3+22 y+3 { y}^2\right) }{2{\left(1- y\right)}^4}\ln^2 y \nonumber \\[1mm]
&  \phantom{xx}  -\frac{ 24 y \left(1+ y\right) }{{\left(1- y\right)}^3}\ln y -\frac{ 94 y}{{\left(1- y \right)}^2}\,. \nonumber 
\end{align}
Here $\HPL(1,0,-1,0;y)$ is a  harmonic polylogarithm of weight four  with two indices different from zero, which we evaluate numerically with the help of the program {\tt HPL} \cite{Maitre:2005uu,Maitre:2007kp}. The polylogarithm of order three (four) is denoted by $\Li_3 (z)$~$\big($$\Li_4 (z)$$\big)$, while  $\zeta_2 = \pi^2/6$, $\zeta_3 \simeq 1.20206$ and $\zeta_4 = \pi^4/90$ are the relevant  Riemann's zeta values. Finally, the functions multiplying the logarithms $\ln \mu_R^2/m_q^2$ in (\ref{eq:digluon}) are given by ${\cal M}_P (\tau) = 2 \hspace{0.25mm} \tau \hspace{0.25mm}  {\cal P}^\prime (\tau)$ and ${\cal M}_S (\tau) = 2 \hspace{0.25mm} \tau \hspace{0.25mm}  {\cal S}^\prime (\tau)$ with the prime denoting a derivative with respect to $\tau$.

In the case of the diphoton partial widths, one has 
\begin{align}
\Delta_{\phi}^\gamma & =  \frac{\alpha_s}{\pi}  \left(  {\cal A}_\phi (y_{\beta_{q/\phi}})+  {\cal M}_\phi (\tau_\phi^q) \hspace{0.25mm} \ln \frac{\mu_q^2}{m_q^2}  \right)\,,
\end{align}
with  \cite{Harlander:2005rq}
\begin{align}
{\cal A}_P (y) & = - \frac{ y \left(1+y^2\right) }{{\left(1-y\right)}^3 (1+y)}  \biggl[ 72\Li_4(y) + 96\Li_4(-y) - \frac{128}{3} \hspace{0.25mm} \big[  \Li_3(y) + \Li_3(-y) \big ] \ln y  \nonumber \\[1mm] 
& \hspace{3.15cm} + \frac{28}{3} \Li_2(y)\ln^2 y + \frac{16}{3} \Li_2(-y)\ln^2 y + \frac{1}{18}\ln^4 y + \frac{8}{3} \hspace{0.25mm}\zeta_2\ln^2 y + \frac{32}{3}\hspace{0.25mm} \zeta_3\ln y + 12\hspace{0.25mm}\zeta_4 \biggr]  \nonumber \\[1mm] 
\\[-5mm]
\nonumber 
\\[-5mm]
& \phantom{xx}  +\frac{y }{{\left(1-y\right)}^2} \biggl[ -\frac{56}{3} \Li_3(y) - \frac{64}{3} \Li_3(-y) + 16 \Li_2 (y) \ln y + \frac{32}{3} \Li_2(-y)\ln y  \nonumber \\[1mm] 
& \hspace{2.2cm} +\frac{20}{3} \ln \left ( 1 - y \right ) \ln^2 y -\frac{8}{3} \hspace{0.25mm}\zeta_2\ln y + \frac{8}{3} \hspace{0.25mm} \zeta_3 \biggr] +\frac{2y \left(1+y\right) } {3 {\left(1-y\right)}^3} \ln^3 y \,, \nonumber  \\[2mm] 
{\cal A}_S (y) & = - \frac{ y \left(1+y+y^2+y^3\right) }{{\left(1-y\right)}^5}  \biggl[ 108\Li_4(y) + 144\Li_4(-y) - 64 \hspace{0.25mm} \big [ \Li_3(y) +\Li_3(-y) \big ] \ln y \nonumber \\[1mm] 
& \hspace{3.75cm} + 14\Li_2(y)\ln^2 y + 8\Li_2(-y)\ln^2 y + \frac{1}{12}\ln^4 y + 4\hspace{0.25mm}\zeta_2\ln^2 y + 16\hspace{0.25mm} \zeta_3\ln y + 18\hspace{0.25mm}\zeta_4 \biggr]  \nonumber \\[1mm] 
& \phantom{xx}  +\frac{y {\left(1+y\right)}^2}{{\left(1-y\right)}^4} \biggl[ -32\Li_3(-y) + 16\Li_2(-y)\ln y  -4\hspace{0.25mm}\zeta_2\ln y \biggr] - \frac{4\, y \left(7-2 y+7 y^2\right) }{{\left(1-y\right)}^4} \Li_3(y) \\[1mm] 
& \phantom{xx} +\frac{8 y \left(3-2 y+3 y^2\right) }{{\left(1-y\right)}^4} \Li_2(y)\ln y +\frac{2y \left(5-6 y+5 y^2\right) } {{\left(1-y\right)}^4} \ln \left (1-y \right )\ln^2 y +\frac{y \left(3+25 y-7 y^2+3 y^3\right) } {3 {\left(1-y\right)}^5} \ln^3 y \nonumber \hspace{6mm} \\[1mm] 
& \phantom{xx} +\frac{4 y \left(1-14 y+y^2\right) } {{\left(1-y\right)}^4} \hspace{0.25mm} \zeta_3 +\frac{12 y^2 } {{\left(1-y\right)}^4} \ln^2 y -\frac{12 y \left(1+y\right) }{{\left(1-y\right)}^3} \ln y -\frac{20 y} {{\left(1-y\right)}^2} \,. \nonumber 
\end{align}

Above the $b \bar b$ threshold, the assumption that the new spin-0 states decay into bottom-quark pairs only through mixing with the corresponding QCD bound states becomes inadequate. Following  \cite{Drees:1989du}, we interpolate between the resonance region and the region where perturbative QCD is applicable  using a heuristic model that is inspired by QCD sum rules. The interpolations take the form 
\begin{align}  
{\cal N}^{b}_P & = 1 - \exp \left [ - 7.39  \left ( 1 - \frac{(m_B + m_{B^\ast})^2}{m_P^2} \right )^{2.5 \,} \right ] \,, \label{eq:BastB} \\[2mm]
{\cal N}^{b}_S & = 1 - \exp \left [ -  8.63 \left ( 1 - \frac{4 m_B^2}{m_S^2}  \right )^{0.8 \,} \right ] \,, \label{eq:BB}
\end{align}
with $m_B = 5.28 \, {\rm GeV}$ and  $m_{B^\ast} = 5.33\, {\rm GeV}$ \cite{Agashe:2014kda}. These expressions update the results obtained previously in \cite{Drees:1989du,Baumgart:2012pj}. In practice, the interpolation is achieved by multiplying the partonic decay widths $\Gamma \hspace{0.25mm} (P,S \to b \bar b)$ by the factors ${\cal N}^{b}_{P,S}$ introduced above. 

The partial decay widths of the spin-0 bottomonium states to gluons are given to LO in $\alpha_s$ and $v_b$ by (see for example \cite{Petrelli:1997ge,Drees:1989du,Baumgart:2012pj})
\begin{align}
\Gamma \hspace{0.25mm} ( \eta_b (n) \to gg ) & = \frac{\alpha_s^2}{3 m_{\eta_b (n)}^2} \, \big |R_{\eta_b (n)} (0) \big  |^2 \,, \\[1mm]
\Gamma \hspace{0.25mm} ( \chi_b (n) \to gg ) & = \frac{3 \alpha_s^2}{m_{\chi_b (n)}^4} \, \big |R^\prime_{\chi_b (n)} (0) \big  |^2 \,.
\end{align}
The partial decay widths to gluons essentially saturate  the total decay widths for $\eta_b (n)$ with $n \neq 5,6$ and all the $\chi_b (n)$ states. In the case of  $\eta_b (5)$ and $\eta_b (6)$, however, also decays to final states involving $\pi$ and $B_{(s)}$ mesons are relevant. We use \cite{Agashe:2014kda}
\begin{align}
\Gamma \hspace{0.25mm} (  \eta_b (5)  & \to \pi \; \text{mesons})  = 1.5 \, {\rm MeV} \,, \\[1mm]
\Gamma \hspace{0.25mm} (  \eta_b (6)  & \to \pi \;  \text{mesons})  = 3  \, {\rm MeV}  \,,
\end{align}
for the decays to pion final states, while the $B_{(s)}$ decays are incorporated via the approximate relations \cite{Baumgart:2012pj}
\begin{align}
& \Gamma \hspace{0.25mm} ( \eta_b (5) \to B +  B_{s} \; \text{mesons})   \simeq 0.9 \hspace{0.5mm} \Gamma \hspace{0.25mm} ( \Upsilon(5)  \to B \; \text{mesons}) +  0.65 \hspace{0.5mm} \Gamma \hspace{0.25mm} ( \Upsilon(5)  \to B_{s} \; \text{mesons}) \,, \\[2mm]
& \hspace{6mm} \Gamma \hspace{0.25mm} ( \eta_b (6) \to B + B_{s} \; \text{mesons})   \simeq  \Gamma \hspace{0.25mm} ( \Upsilon(5)  \to B \; \text{mesons}) +  \Gamma \hspace{0.25mm} ( \Upsilon(5)  \to B_{s} \; \text{mesons}) \,, 
\end{align}
where \cite{Agashe:2014kda}
\begin{align}
\Gamma \hspace{0.25mm} (  \Upsilon(5)  \to B \; \text{mesons} )  & = 42 \, {\rm MeV} \,, \\[1mm]
\Gamma \hspace{0.25mm} (  \Upsilon(5)  \to B_s \; \text{mesons} )  & = 11 \, {\rm MeV}  \,.
\end{align}

In order to make our discussion of  dimuon branching ratios of spin-0 bottomonium states self-contained, we collect below the  formulas used in Sec.~\ref{sec:etabdimuonBr}.  The dimuon decays of the $\eta_b (n)$ and  $\chi_b (n)$ mesons proceed at lowest order in QED through one-loop diagrams involving a two-photon intermediate state. The imaginary (absorptive) parts of the corresponding amplitudes can be calculated in a model-independent fashion using unitarity arguments \cite{Quigg:1968zz}. Such an approach leads to the following unitarity bounds on the dimuon branching ratios in question (see for instance~\cite{Martin:1970ai})
\begin{align}
{\rm Br} \hspace{0.25mm} \big (\eta_b (n) \to \gamma^\ast \gamma^\ast \to \mu^+ \mu^- \big ) & \geq \frac{\alpha^4}{36 \alpha_s^2} \, \frac{m_\mu^2}{m_{\eta_b (n)}^2} \, \frac{1}{\beta_{\mu/\eta_b (n)}} \, \ln^2 x_{\beta_{\mu/\eta_b (n)}}  \,, \\[2mm]
{\rm Br} \hspace{0.25mm} \big (\chi_b (n) \to \gamma^\ast \gamma^\ast \to \mu^+ \mu^- \big ) & \geq \frac{\alpha^4}{9 \alpha_s^2} \,  \frac{m_\mu^2}{m_{\eta_b (n)}^2}  \, \beta_{\mu/\chi_b (n)} \, \ln^2 x_{\beta_{\mu/\chi_b (n)}}  \,,
\end{align}
with $m_\mu = 0.106 \, {\rm GeV}$ the muon mass. These results assume that the relevant branching ratio to two gluons is close to 1, which is a very good approximation for all spin-0 bottomonium states, but~$\eta_b(5)$ and~$\eta_b(6)$. In contrast to the absorptive parts of the amplitudes, their dispersive (real) parts depend on how the~$\eta_b(n)\gamma\gamma$ and~$\chi_b(n)\gamma\gamma$ form factors are modeled. Using a constituent quark model \cite{Ametller:1983ec}, we find that including real parts into the calculations lead to branching ratios that are only slightly above the unitarity bounds. To get an order of magnitude estimate of the QED contributions to the dimuon branching ratios of spin-0 bottomonium states the above formulas are hence sufficient. 

Besides QED contributions, the dimuon rates of the  $\eta_b (n)$ and  $\chi_b (n)$ mesons also receive purely EW corrections associated to virtual $Z$-boson and Higgs exchange. Assuming again that the branching ratios to gluon pairs fully dominate, we obtain 
\begin{align}
{\rm Br} \hspace{0.25mm} \big (\eta_b (n) \to Z^\ast\to \mu^+ \mu^- \big ) & = \frac{9 \hspace{0.25mm} G_F^2}{16 \pi^2 \alpha_s^2} \, m_\mu^2 \hspace{0.5mm}  m_{\eta_b (n)}^2 \, \beta_{\mu/\eta_b (n)} \,, \\[2mm]
{\rm Br} \hspace{0.25mm} \big (\chi_b (n) \to h^\ast \to \mu^+ \mu^- \big ) & =  \frac{9 \hspace{0.25mm} G_F^2}{\pi^2 \alpha_s^2} \, m_\mu^2 \hspace{0.5mm} m_b^2 \, \beta_{\mu/\chi_b (n)}^3  \,  \frac{m_{\chi_b (n)}^4}{m_h^4} \,.
\end{align}
Here $G_F = 1.167 \cdot 10^{-5} \, {\rm GeV}^{-2}$ denotes the Fermi constant. The latter formulas can be shown to agree with results obtained  in the context of $\pi^0$ and $K^0$ decays (cf.~\cite{Kalogeropoulos:1979wv,Ecker:1991ru}). 

\end{widetext}

\end{appendix}

\end{document}